\documentclass[aps,prb,10pt,twocolumn,superscriptaddress, longbibliography]{revtex4-1}
\usepackage{amsmath}
\usepackage{amsfonts}
\usepackage{amssymb}
\usepackage{graphicx}
\usepackage{bm}
\usepackage[colorlinks=true,linktocpage=true,breaklinks=true,
allcolors=blue]{hyperref}
\usepackage{xcolor}

\begin{document}

\title{Microscopic origin of scalar potential induced topological transition \\in massive Dirac fermions and scalar Hall effect}

\author{Sumit Ghosh}
\email{s.ghosh@fz-juelich.de}
\affiliation{Max Born Institute for Nonlinear Optics and Short Pulse Spectroscopy, 12489, Berlin, Germany}
\affiliation{Peter Gr{\"u}nberg Institut (PGI-1), Forschungszentrum J{\"u}lich GmbH, 52428 J{\"u}lich, Germany}
\affiliation{Institute of Physics, Johannes Gutenberg-University Mainz, 55128 Mainz, Germany}

\author{Yuriy Mokrousov}
\affiliation{Peter Gr{\"u}nberg Institut (PGI-1), Forschungszentrum J{\"u}lich GmbH, 52428 J{\"u}lich, Germany}
\affiliation{Institute of Physics, Johannes Gutenberg-University Mainz, 55128 Mainz, Germany}

\author{Stefan Bl\"ugel}
\affiliation{Peter Gr{\"u}nberg Institut (PGI-1), Forschungszentrum J{\"u}lich GmbH, 52428 J{\"u}lich, Germany}

\begin{abstract}
We present a systematic study of scalar potential induced topological transition in massive Dirac fermions. We show  how a distribution of scalar potential can manipulate the signature of the gap or the \textit{mass}, as well as the dispersion leading to a band inversion. This is mediated by the \textit{Klein tunnelling} as well as  \textit{inverse Klein tunnelling} which makes it inherently different from the mechanism leading to topological Anderson insulator. In one dimension it can lead to the formation of edge localisation. In two dimensions this can give rise to the quantised Hall effect. Unlike conventional Hall effects, this  is induced by a scalar interaction and intrinsic in nature. Therefore we call it a \textit{scalar Hall effect}. This can facilitate a direct manipulation of topological invariants, e.g. the Chern number, as well as the manipulation of the edge states locally in a trivial insulator and thus opens new possibilities for tuning physical observables which originate from the nontrivial topology.
\end{abstract}

\maketitle

\section{Introduction}
In recent years \textit{topology} has become a central concept in condensed matter physics \cite{Ghosh2023}, while materials with non-trivial topological properties have become a key ingredient in designing next-generation transport and memory devices~\cite{Breunig2021, Gilbert2021}. An immense effort has been directed at discovering suitable materials \cite{Lin2015, Ando2013} and characterization of their topological classes \cite{Altland1997, Ryu2010}. This highly active research field has brought several new topological phases in the last decade \cite{Armitage2018, Lv2021}. The first topological insulator, namely the Quantum Hall Insulator, was discovered under a strong magnetic field \cite{Klitzing1980} which is quite challenging for any practical purpose. In materials with strong spin-orbit-coupling (SOC), such as HgTe \cite{Bernevig2006} or Bi$_2$Se$_3$ \cite{Zhang2009}, the nontrivial topological features arise from electronic interactions involving orbital and spin degrees of freedom. Such interactions can give rise to different topological phases such as the Quantum Anomalous Hall Insulator \cite{Nagaosa2010} and Quantum Spin Hall Insulator \cite{Sinova2015}. Such phases can host \textit{dissipationless} current \cite{Murakami2003} which opens immense possibilities in device applications due to their robustness against scattering. Impurity scattering can also contribute to the Hall current \cite{Fukazawa2017, Crepieux2001, Cullen2023} via mechanisms like side jump and skew scattering which is known as the \textit{extrinsic} contribution. This is inherently different from the \textit{intrinsic} contribution arising from the internal characteristics of the systems which is often described in terms of the topology of the system \cite{Thouless1982}. Several physical observables sharing connection to real \cite{Ghosh2022} and reciprocal \cite{Ghosh2017} space topology are known to be enhanced in the presence of scalar impurity which indicates that scalar potential might have a deeper connection to the intrinsic contribution as well. A proper theoretical description of such a connection is still missing.

In certain cases it is possible to tune the topological properties with magnetic impurities \cite{Cheng2021, Shamim2021}, however, these cases are very material specific. For most of the cases, SOC is considered to be the source of nontrivial topological properties which manifest itself via band inversion \cite{Zhu2012}. The band inversion is indeed an impeccable sign of a topological transition, however, on its own, it does not directly reflect the specific mechanism which drives it. Besides, it is also possible to have a topological insulator even without SOC or magnetic moment \cite{Haldane1988, Fu2011} by exploiting the symmetry of electronic degrees of freedom. A proper description of the mechanism behind inducing topological features and the transition dynamics between different topological phases is therefore highly desired to access and manipulate the topological phases of solid-state systems.

The key to accessing the non-trivial topological properties of these systems lies in their dispersion which resembles that of a relativistic particle. One of the characteristic features of a relativistic dispersion is that each gap is associated with a well-defined \textit{signature}. For any generic Dirac spinor $\psi$ obeying $( i\gamma^{\mu} \partial_{\mu} - m)\psi$=0 (in natural units $c$=$\hbar$=1), where $\gamma^{\mu}$ ($\mu$=0 being the temporal component) are the Dirac matrices obeying the Clifford algebra $\lbrace \gamma^\mu, \gamma^\nu \rbrace$=$2 \eta^{\mu \nu} \mathbb{I}$ with $\eta$ being the Minkowski metric, the spectrum has a fundamental gap of $2m$. This is commonly known as the \textit{mass-gap} since the gap is associated with the rest mass of the particle in relativistic theory. Each energy band is associated with a particular sign of $\langle \gamma^0 \rangle$ and as a result, each gap can be identified with a specific signature.  While the eigenvalue spectra remain the same irrespective of the sign of $m$, the eigenstates are sensitive to it and manifest different topological features. A pronounced example is the appearance of Jackiw-Rabi modes at the boundary of two different domains characterised by opposite mass terms \cite{Jackiw1976} where the topological phase boundary is manifested as spatial localisation which is the essence of edge states in topological insulators. 

To understand the relation between the mass term and topological characteristics of a system, let us consider a generic $2 \times 2$ Dirac Hamiltonian $\bm{\sigma}\cdot \bm{n}(\bm{R})$ defined on Bloch sphere, where $\bm{\sigma}$ is the vector of Pauli matrices and $\bm{n}(\bm{R})$ is the unit vector parameterised by $\bm{R}$. In this case, the vector Berry curvature is simply given by $\bm{n}$ \cite{Xiao2010}. In two dimensions, two of the Pauli matrices are coupled with momentum corresponding to the direction of motion while the third is coupled to the mass term. The Chern number in this case follows the same signature of the out-of-plane component of $\bm{n}$, which is nothing but the mass term. A similar correlation has been observed in two-dimensional paramagnetic systems and three-dimensional complex heterostructures as well \cite{Ghosh2019}. In a condensed matter system, the mass term is associated with an order parameter and therefore can be exploited to identify different phases \cite{Ghosh2016}. 
In a real system, however, manipulating the mass term is quite nontrivial. The mass term does not commute with the rest of the Hamiltonian and thus results in intriguing new topological features \cite{Kane2005,*Kane2005a}. Physically, such non-commuting nature originates from a complex mixture of different degrees of freedom which makes their control quite challenging. A simple way to manipulate the magnitude and signature of the mass gap thus has an enormous potential in fabricating topologically non-trivial systems and exploring their applications.

In this work, taking a generic two-band system as a prototype, we present a systematic analysis of the topological properties in a multi-band system and demonstrate a simplified way to generate and manipulate non-trivial topological features with the help of scalar potential. In practice, such scalar potential can be introduced by the means of nano-patterning \cite{Barad2021} or surface super lattice \cite{Esaki1970}. By using the two-band Dirac Hamiltonian, we demonstrate how one can manipulate the mixture of different quantum states which in turn controls the topological features.  Since the scalar potential is represented by an identity matrix which commutes with any other matrix, our formalism is equally applicable to systems where the non-trivial topology arises from spin or orbital degrees of freedom. The formalism is, thus, applicable to a large class of condensed matter systems, which facilitates a wide range of applications of this generic protocol.


\section{Dirac equation in one dimension} 

To understand the connection between the scalar potential and the mass term, let us first consider the one-dimensional Dirac equation in the continuum limit. In one dimension, it is sufficient to consider the $2 \times 2$ representation of Dirac matrices, which we choose here as the Pauli matrices ($\bm{\sigma}$). We define our system with the one-dimensional Dirac Hamiltonian
\begin{eqnarray}
 H_1^D = -i \sigma_1 \partial_x + \sigma_3 m + \sigma_0 V(x).
 \label{H1}
\end{eqnarray}
where $V(x)$ is a scalar potential $\sigma_0$ is the identity matrix of rank 2. In the absence of the potential term, the energy spectrum consists of two hyperbolic branches ($E=\pm\sqrt{p^2 + m^2}$, $p$ is the momentum) separated by a gap of $2m$ with positive and negative energy eigenvalues characterised by positive and negative values of $\langle \sigma_3 \rangle$. For such a system, it is possible to achieve complete transmission if the barrier height is greater than twice the mass term ($V_0>2m$) (Appendix\,\ref{KT}). This is known as the \textit{Klein paradox} \cite{Klein1929} which has attracted a lot of interest in both high energy physics as well as in condensed matter physics \cite{Katsnelson2006}. The simplest way to understand the underlying mechanism is via the mixture of states with positive and negative energy. If the scalar potential is strong enough ($V_0>2m$), then it can elevate the negative energy states inside the potential barrier to an energy level occupied by the positive energy states outside the barrier which creates a continuous channel using plane-wave modes. For smaller barrier width, interference due to finite size effect is more prominent which is manifested as oscillations in transmission probability (Appendix\,\ref{KT}).

\begin{figure}[t!]
\centering
\includegraphics[width=0.40\textwidth]{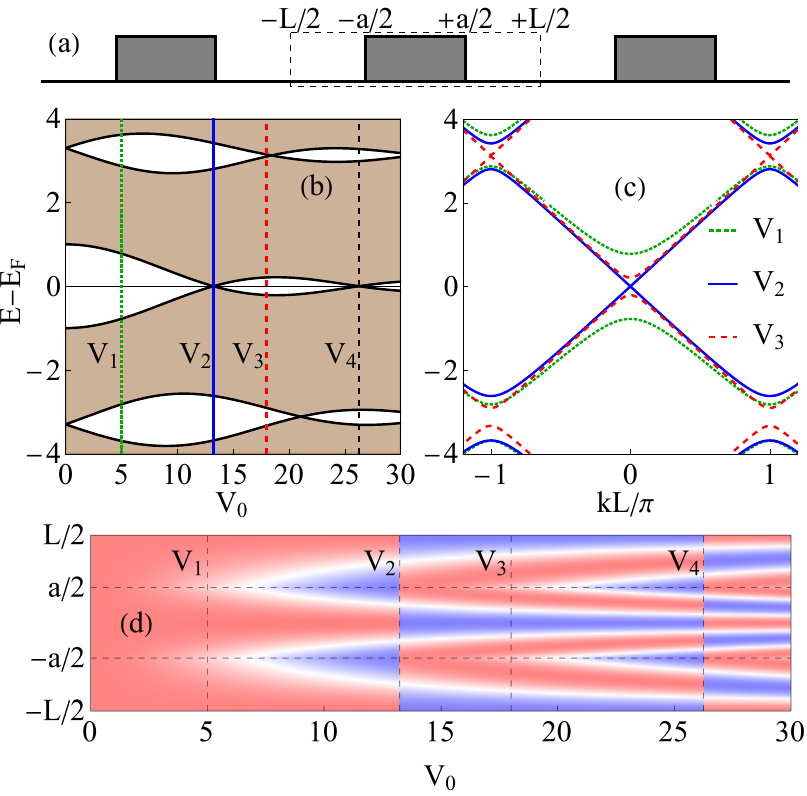}
\caption{Variation of mass-gap with $V_0$. (a) Schematic of the one-dimensional periodic lattice. The dashed box shows the unit cell and the grey boxes show the scalar potential $V_0$. (b) Variation of bandwidth (filled region) with the applied potential. White regions correspond to the band gap. (c) Band structure for selected values of $V_0$ marked by vertical coloured lines in (b). (d) Distribution of the order parameter $\Delta(x)$ over the unit cell with respect to $V_0$ where red and blue colours represent the positive and negative magnitude respectively.}
\label{fig:band1c}
\end{figure}

The physics becomes more intriguing if such potential region is employed periodically which can create states with alternating signs of the mass term in successive regions in space. As a result, although the scalar potential itself cannot alter the mass gap, the interference between states with different mass terms can alter the characteristic band gap. To demonstrate this effect we consider the one-dimensional Dirac-Kronig-Penney model which has been used to analyse relativistic quarks \cite{McKellar1987} and fermions \cite{Ghosh2014}. Here we consider a one-dimensional lattice with length $L$ (set to be 1) and with a rectangular potential of height $V_0$ and width $a$ such that $a/L=0.4$ and calculate the band structure (Fig.\,\ref{fig:band1c}). We define the spatial order parameter 
\begin{eqnarray}
\Delta(x) = \left[ (\langle +, x| \sigma_3 | +, x \rangle - \langle -,x | \sigma_3 | -,x \rangle)/2 \right]_{k=0}
\label{op}
\end{eqnarray}
where $|+,x\rangle$ and $|-,x\rangle$ denote the wave function at $x$ of the lowest positive energy and highest negative energy states. For $V_0$=0 these two states reside at energy $E$=$\pm m$. For $V_0>$0, both of these states are shifted by a positive value. To keep these two states symmetric around the zero level we subtract a fixed energy $E_F$ for each value of $V_0$. This energy is analogous to the Fermi level in a system with a finite number of states\footnote{In the continuum limit, a filled Dirac sea contains an infinite number of particles and the notion of Fermi level is not well defined.}.

\begin{figure}[ht!]
\centering
\includegraphics[width=0.45\textwidth]{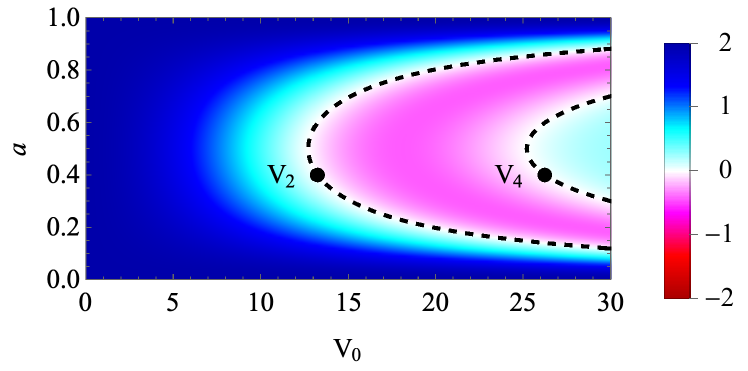}
\caption{Variation of the mass gap with the barrier width ($a$) and barrier height ($V_0$). The colour shows the value of the mass gap multiplied by the sign of the order parameter at $x$=0 and the black dashed lines show the transition boundary for the mass term. $V_2$ and $V_4$ shows the location of transition point in Fig.\,\ref{fig:band1c}.}
\label{fig:av}
\end{figure}

From Fig.\,\ref{fig:band1c} one can see that by increasing the barrier height it is possible to manipulate the mass gap. The mass gap decreases because the potential $V_0$ now pulls up negative energy states within the window $-m<E<m$ which can now tunnel through the region without potential where the evanescent modes have opposite signatures of the mass term. In a sense, this is the \textit{Inverse Klein Tunnelling} where the tunnelling happens through the potential free region. This mechanism promotes more mixing of quantum states which creates a spatial modulation of the order parameter $\Delta$. Note that at some critical values, the gap vanishes completely ($V_2$, $V_4$ in Fig.\,\ref{fig:band1c}), and the distribution of the order parameter also flips sign. This critical potential is minimal when the barrier width is half of the unit cell which maximises the mixture of quantum states (Fig.\,\ref{fig:av}). The flipping of the order parameter establishes that each of these crossings is associated with a band inversion. The impact of the band inversion will become more clear when we explore the topological properties of a two-dimensional model in the next section.

\section{Lattice model: From 1D to 2D} 

The transition from the continuum to a lattice model for a relativistic Hamiltonian is not a straightforward task. Since we are not looking into chiral fermions here, we are free from the obstacles imposed by the Nielson-Ninomiya theorem \cite{Nielsen1981}. For modelling massive/gapped Fermions, one can start from a chiral Fermion and simply add a \textit{mass} term. This produces a pair of Fermions with opposite group velocities within the Brillouin zone, commonly known as \textit{doublers}. To avoid this, one can adopt Wilson's prescription \cite{Wilson1974} and introduce a momentum-dependent coupling between the spinor component. This is also known as \textit{Creutz lattice} \cite{Creutz1994}.  
For a two component spinor field $\psi= [a,b]^T$ the Creutz Hamiltonian can be expressed as 
\begin{eqnarray}
H_{C} &=& iA\sum_{j} a_{j+1}^{\dagger}a_{j} - a_{j}^{\dagger}a_{j+1} - b_{j+1}^{\dagger}b_{j} + b_{j}^{\dagger}b_{j+1} \nonumber \\
&-&B \sum_{j}a_{j+1}^{\dagger}b_{j} + b_{j+1}^{\dagger}a_{j} +  M \sum_{j}a_{j}^{\dagger}b_{j} + b_{j}^{\dagger}a_{j}.
\label{HC}
\end{eqnarray}
For convenience we perform a unitary transformation $H_1 = U^\dagger H_{C} U$ [where $U$=$(\sigma_1+\sigma_3)/\sqrt{2}$]. The transformed Hamiltonian in reciprocal space is given by $H_1(k)$=$[M-2B\cos(k)]\sigma_3 + 2A \sin(k) \sigma_1$. Although the physical outcome does not change under such unitary transformation, the modified form makes it easier to correlate our prediction with known physical systems which would be more clear when we discuss the scenario in two dimensions.

\subsection{Tuning topological phase in 1D: Emergence of edge localisation}

We start from the lattice Hamiltonian in one dimension given by

\begin{eqnarray}
H_1(k)=(M-2B\cos(k))\sigma_3 + 2A \sin(k) \sigma_1
\label{H1l}
\end{eqnarray}

One can readily see that for $k \to 0$, the low energy spectrum corresponds to the continuum Dirac Hamiltonian (Eq.~\ref{H1}) with a mass gap of $M-2B$. Here we choose $M=A=5.0$ and $B=2.0$, and a unit cell with $L=40$ sites. A scalar potential of strength $V_0$ is spanned over a region of $a=16$ sites. In the absence of any scalar potential, the system has a \textit{mass} $M-2B$=1 which is expected in any massive Dirac system. The Fermi level corresponds to half-filling and is kept at 0. We define a site-resolved order parameter
\begin{eqnarray}
\Delta_i = \left[ (\langle +, i| \sigma_3 | +, i \rangle - \langle -,i | \sigma_3 | -,i \rangle)/2 \right]_{k=0},
\label{op1}
\end{eqnarray}
where $|+,i\rangle$ and $|-,i\rangle$ is the wave function at site $i$ of the lowest positive energy and highest negative energy states.

\begin{figure}[h!]
\centering
\includegraphics[width=0.48\textwidth]{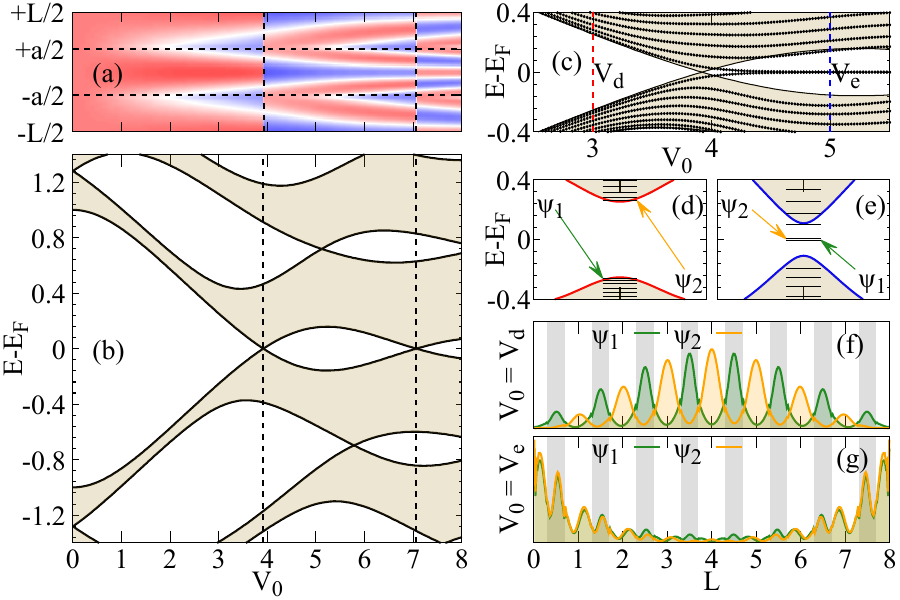}
\caption{Tuning the phases of a 1D chain. (a) Variation of site-resolved order parameter $\Delta_i$ with $V_0$ in a one-dimensional model. (b) Variation of band gap with $V_0$. (c) Variation of energy eigenvalues of a finite chain (dotted lines). The shaded region shows the energy range spanned by a periodic chain. Vertical red and blue dashed lines show two specific values of $V_0$ ($V_d$=3 and $V_e$=5 respectively). (d) and (e) show the band structure close to $k$=0 for $V_0$=$V_d$ and $V_0$=$V_e$ marked with red and blue dashed lines in (c). The horizontal black lines show the energy levels of a finite system. (e) and (f) show the probability density of the highest occupied ($\psi_1$) and lowest unoccupied ($\psi_2$) states marked with orange and green arrows in (d) and (e) respectively. $L$ denotes the length in units of the unit cell and the grey regions denote the potentials.}
\label{fig:band1l}
\end{figure}

From Fig.\,\ref{fig:band1l}a, one can readily see that the mass term behaves similarly compared to the continuum model with respect to $V_0$. The variation of the band gap in the lattice model (Fig.\,\ref{fig:band1l}b) is also qualitatively the same as the prediction of the continuum model. This establishes the validity of the lattice model for our study. From the band structure, one can see that the distribution of the mass term switches signs as it passes through a band crossing (denoted by vertical black dashed lines in Fig.\,\ref{fig:band1l}a,b). To understand if these jumps are associated with any change in the topological phase, we consider a supercell with 8 unit cells (total 320 sites) and with open boundary conditions. We consider two different strengths of the scalar potential ($V_d$ and $V_e$ in Fig.\,\ref{fig:band1l}c) on either side of the crossing point. One can readily see that after the critical potential, the finite chain has eigenvalues close to zero energy (Fig.\,\ref{fig:band1l}c). The eigenstates corresponding to these zero energy modes are strongly localised near the edges (Fig.\,\ref{fig:band1l}g) whereas before transition the highest occupied and lowest unoccupied states are localised in the bulk (Fig.\,\ref{fig:band1l}f). One can see that the system is essentially behaving like an SSH model \cite{Su1979} where the variation of the hopping parameter can be achieved with the scalar potential.  This indicates that one can tune the \textit{mass} of the system with a scalar potential which in turn can influence the topology of the system. This connection will be more clear in the next section where we discuss the two-dimensional systems.

\subsection{Tuning topological phase in 2D: Emergence of quantised Hall effect}

The extension of Hamiltonian $H_1$ (Eq.~\ref{H1l}) to two dimensions is quite straightforward and is given by
\begin{eqnarray}
H_2(k_x,k_y) &=& [M-2B(\cos(k_x)+\cos(k_y))]\sigma_3 \nonumber \\
&& + 2A[\sin(k_x)\sigma_1 + \sin(k_y)\sigma_2].
\label{BHZ}
\end{eqnarray}

\begin{figure}[tp!]
\centering
\includegraphics[width=0.48\textwidth]{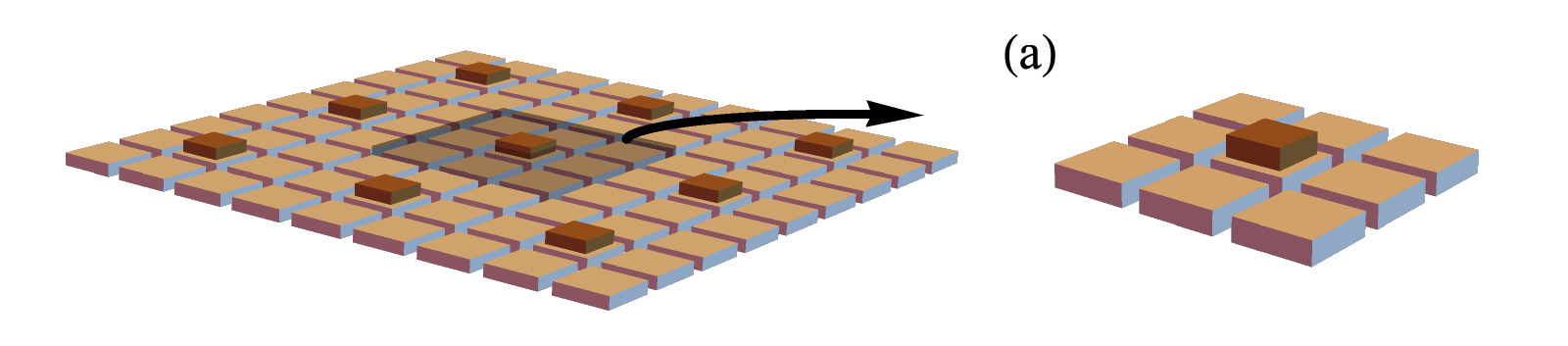}
\includegraphics[width=0.48\textwidth]{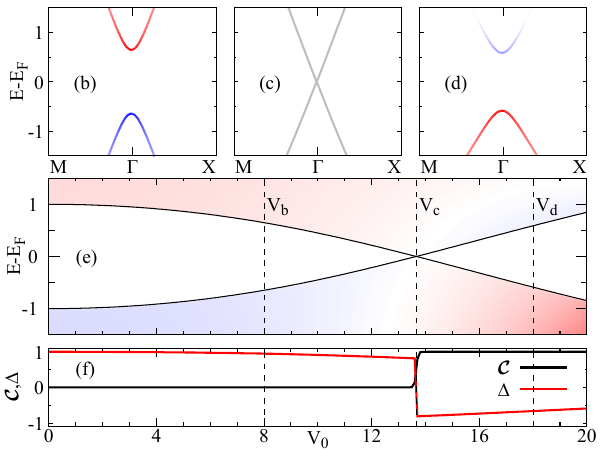}
\caption{Variation of topological nature with $V_0$. (a) Schematic of the supercell lattice with $3\times 3$ motif. Dark brown blocks correspond to areas with an additional scalar potential. (b-d) show the band structure for $V_0=V_b, V_c, V_d$, respectively (shown by the vertical dashed line in (e)) with red (blue) colour showing positive/negative values of $\langle \sigma_3 \rangle$. (e) shows the variation of the band gap with $V_0$ where red (blue) colour corresponds positive(negative) values of $m(E)$. (f) shows the variation of the Chern number (black) and the total order parameter $\Delta$ (red) with $V_0$.}
\label{fig:band2}
\end{figure}

One can easily recognise that this is the $2 \times 2$ block of the Hamiltonian used to define the quantum spin Hall effect in HgTe-CdTe quantum wells, commonly known as the Bernevig-Hughes-Zhang (BHZ) model \cite{Bernevig2006}.
The Creutz lattice (Eq.\ref{HC}) and the BHZ model (Eq.\ref{BHZ}) are the simplest models of a Chern insulator. In the following sections, we use the BHZ model which exhibits a Chern number $\mathcal{C}=sgn(M)$ for $|M|<|4B|$ and $\mathcal{C}=0$ for $|M|>|4B|$. This lends us a perfect playground for further predictions. We start with $M=9$, $B=2$ and $A=5$ which gives a  trivial Chern insulator state ($\mathcal{C}=0$). The mass gap at $k_x,k_y=0$ is given by $M-4B$ which for our choice of parameters is 1. Here we consider a $3\times 3$ supercell with one scalar potential (resulting in 11.1\% coverage) and calculate the variation of band structure, mass term and the Chern number (Fig.\,\ref{fig:band2}) with the variation of $V_0$. 
The Chern number can be calculated from
\begin{eqnarray}
\mathcal{C} &=& \frac{1}{2\pi}\sum_n^{n_F}\int_{BZ} dk_x dk_y [f(E_n)-f(E_m)] \times \nonumber \\
&& \sum_{m \neq n}2 ~{\rm Im} \left[ \frac{\langle m | \partial H/ \partial k_y | n \rangle \langle n | \partial H/ \partial k_x | m \rangle }{(E_n-E_m)^2} \right]
\end{eqnarray}
where $n_F$ is the number of states below the Fermi level. $f(E_n)$ is the Fermi-Dirac distribution for the $n$th energy eigenvalue and $|n\rangle$ is the $n$th eigenstate. The mass term at any particular energy is given by
\begin{eqnarray}
m(E)=\int_{BZ} dk_x dk_y \sum_n^{2N} \langle n | \sigma_3 \otimes \mathbb{I}_N | n \rangle \delta(E_n-E),
\end{eqnarray}
where $N$ is the number of sites (which for our case is 9) and $\mathbb{I}_N$ is the identity matrix of rank $N$. $\delta(x)$ is the Dirac delta function which is approximated as a Lorentzian with broadening $\eta$ which we choose to be 0.005. For a better comparison, we use the same broadening in the calculation of the Chern number as well. Here we use the integrated order parameter defined as
\begin{eqnarray}
\Delta = [(\langle + |\sigma_3 \otimes \mathbb{I}_N | + \rangle - \langle - |\sigma_3 \otimes \mathbb{I}_N | - \rangle)/2]_{k=0}.
\end{eqnarray}
which is sufficient to denote the phase transition.

\begin{figure}[ht!]
\centering
\includegraphics[width=0.48\textwidth]{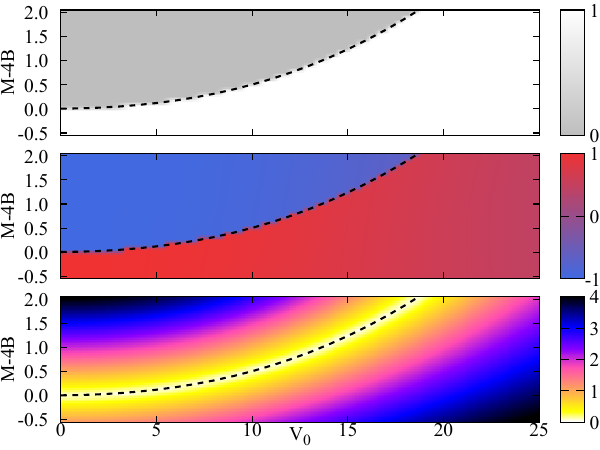}
\caption{Variation of (a) Chern number ($\mathcal{C}$), (b) order parameter ($\Delta$) and (c) mass gap with the parameter $M$ and strength of scalar potential ($V_0$). The black dashed line shows the transition boundary.}
\label{fig:mv}
\end{figure}

\begin{figure}[ht!]
\centering
\includegraphics[width=0.48\textwidth]{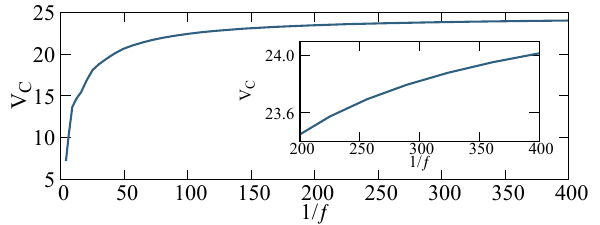}
\caption{Variation of critical potential $V_C$ with the density of scalar potential $f$.}
\label{fig:vc}
\end{figure}

With these definitions, one can clearly see the connection between the band inversion and the \textit{mass} (Fig.\,\ref{fig:band2}) in two dimensions as well. From Fig.\,\ref{fig:band2}, one can see that with the increase of the scalar potential, there is a band gap closing and reopening, similar to what we observe in the one-dimensional case (Fig.\,\ref{fig:band1c},\ref{fig:band1l}). At the critical point, where the bands touch each other, the order parameter as well as the Chern number undergo a jump indicating a change in the topological phase. This is consistent with our earlier picture of band inversion through the mixing of different mass regimes. The non-zero Chern number indicates the generation of a Hall current due to the scalar potential. Since the effect is triggered by a scalar term rather than a vector field, we call it \textit{scalar Hall Effect}. This is to distinguish its origin. In spirit it is an emergent quantum anomalous Hall effect caused by the mixture of quantum states mediated by the scalar potential. In a topologically trivial regime, each occupied band is comprised of states with the same sign of the mass term. As we increase the potential, the interchange of states with opposite mass terms takes place. An increase of the scalar potential enhances the mixing of states with opposite mass terms and thus transports the system from a topologically trivial to a non-trivial phase, characterised by a non-zero Chern number. The critical potential ($V_C$ in Fig.\,\ref{fig:band2}) at which the transition takes place shows a parabolic dependence with respect to the parameter $M$ that controls the \textit{mass} of the system (Fig.\,\ref{fig:mv}) and a logarithmic behaviour with respect to the potential concentration (Fig.\,\ref{fig:vc}).

There is an important difference between the underlying mechanism leading to scalar Hall effect and that leading to a Topological Anderson Insulator \cite{Groth2009,*Li2009}. Theory on Anderson insulator shows that the topological phase is formed by the dissipative state whereas the scalar Hall effect is caused by plane waves mediated through (inverse)Klein tunnelling. While topological Anderson phase is observed within initial metallic regime, scalar Hall effect emerges in the gapped region. In contrast to random disorder-induced Chern insulator \cite{Kuno2019}, this mechanism is intrinsic in nature. This is reflected in the fact that the emergent non-trivial phase possesses the same magnitude of the gap compared to the potential free case whereas the disorder-induced non-trivial phase is known to have an order of magnitude smaller gap \cite{Kuno2019}. The underlying mechanism is also distinct from the previously reported mechanism for voltage-modulated Chern number \cite{Jiang2010} where one can arbitrarily enhance the Chern number by increasing the voltage. Our mechanism, on the other hand, facilitates a transition from $\mathcal{C}$=0 to $\mathcal{C}$=1, which is the highest Chern number possible for this model. If one starts from a topologically non-trivial configuration, the additional scalar potential enhances the topological protection and reduces the mixing of quantum states and thus prevents any further topological transition (Fig.\,\ref{fig:mv}).


\section{Formation and manipulation of edge states} 

\begin{figure}[ht!]
\centering
\includegraphics[width=0.48\textwidth]{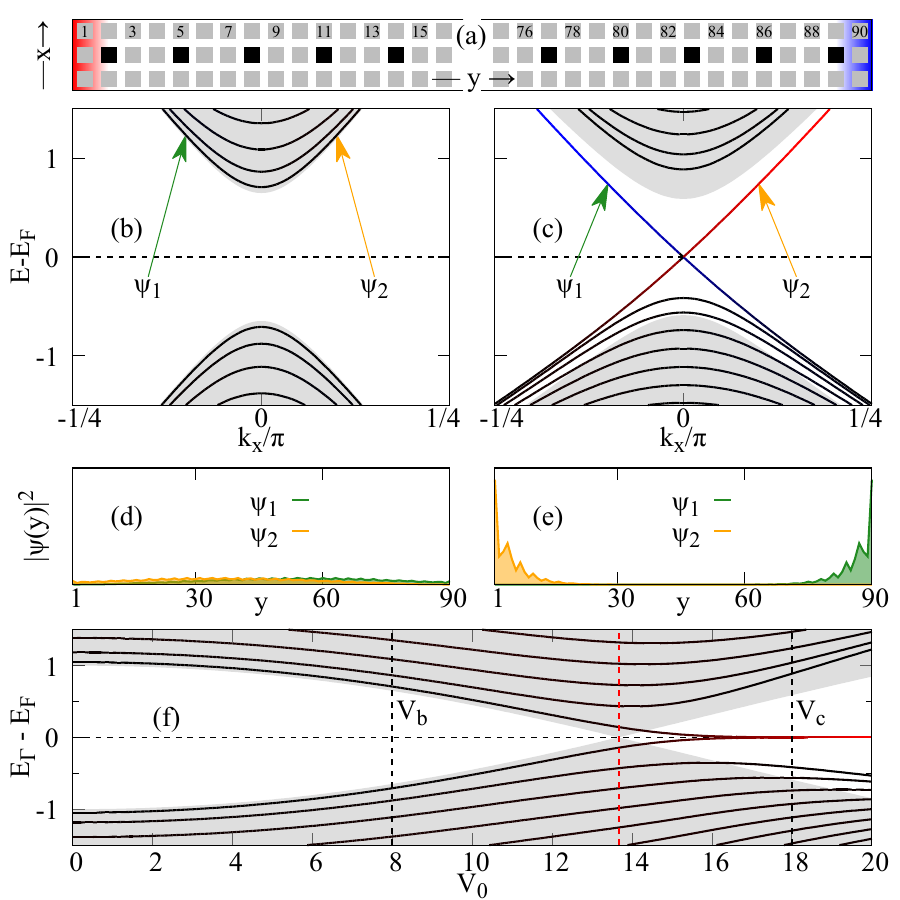}
\caption{Variation of bulk and edge properties with $V_0$. (a) The schematic of the ribbon with scalar potentials is denoted in black. (b) and (c) show the band structure for $V_0$=$V_b$ and $V_0$=$V_c$ ($V_b$ and $V_d$ in Fig.\,\ref{fig:band2}e). Red, blue and black colours mark the contribution from the edge at $y$=1, the edge at $y$=90 and the bulk region. Gray shaded region marks the region spanned by the bulk bands. (d),(e) show the probability density of the state $\Psi_{1,2}$, marked with green and orange arrow in (a),(b). (c) shows the variation of eigenvalues at the $\Gamma$-point where the black and red colour correspond to the contribution from the bulk and edges. The grey region shows the area spanned by the bulk bands and the vertical red dashed line shows the point of band inversion ($V_0$=$V_c$ in Fig.\,\ref{fig:band2}c,e,f).}
\label{fig:band2e}
\end{figure}

A non-vanishing bulk topological invariant shares a direct correspondence with the existence of edge states \cite{Hatsugai1993, Qi2006}. To demonstrate that, we consider a ribbon configuration, i.e. we assume a periodic boundary condition along $x$-direction, and open boundary condition along $y$-direction by repeating the block shown in Fig.\,\ref{fig:band2}a. Here we choose 3 sites along $x$ and 90 sites along $y$ (total 270 sites, Fig.\,\ref{fig:band2e}) and introduce scalar potential $V_0$ in one in every nine ($3\times 3$) sites ($11.1\%$ coverage).

 In such a ribbon the edge states emerge when $V_0$ crosses the critical value at which the bulk bands cross each other (Fig.\,\ref{fig:band2e}). This is similar to the emergence of edge localisation in a 1D system (Fig.\,\ref{H1l}). Each edge hosts a single edge state such that opposite edges host states with an opposite group velocity, which is expected in the case of a Chern insulator with a Chern number of 1.

\begin{figure}[h!]
\centering
\includegraphics[width=0.48\textwidth]{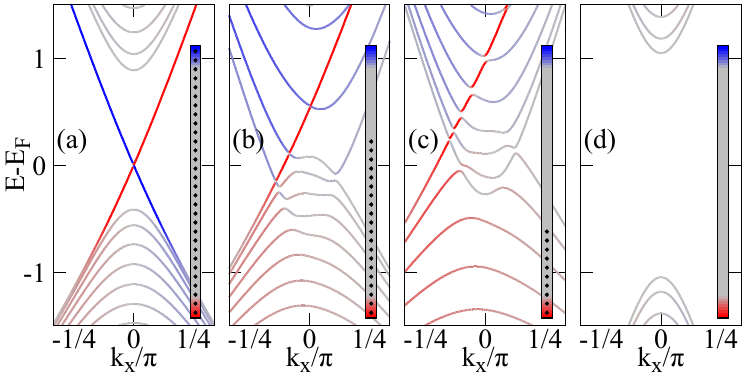}
\caption{Variation of the band structure with the distribution of the scalar potential along the ribbon. (a) The band structure when the full width of the ribbon is covered with the potential. (b,c) show the band structure with 2/3 and 1/3 coverage, and (d) shows the band structure with no potential. Inset of each figure shows the spatial distribution where the black dots represent the scalar potential. Red and blue colours denote the $y$=1 and $y$=90 edges (Fig.\,\ref{fig:band2e}a).}
\label{fig:rib}
\end{figure}

One can further manipulate the behaviour of the edge states locally by controlling the distribution of the scalar potential. As we explained before, the scalar potential enhances the mixing between the states with an opposite mass term which causes the topological transition. In an extended system, one can use the scalar potential selectively in different regions of space to infuse the topological nature selectively. To demonstrate this we consider the aforementioned ribbon (Fig.\,\ref{fig:band2e}). Then we start removing the scalar potential from one end and calculate the band structure (Fig.\,\ref{fig:rib}). The Fermi level ($E_F$) is defined as the middle of 270$th$ and 271$th$ eigenvalue at the $\Gamma$-point.

With this simple procedure, one can easily manipulate the edge states selectively. By removing the scalar potential at one edge, we reduce the mixing of the states locally and as a result, the states which were sharply localised at the edges before now start moving more into the central region. This is manifested by the fact that the sharp red line in Fig.\,\ref{fig:rib} remains intact as long as there are scalar potentials at the corresponding edges whereas the blue lines fade out and mix strongly with the grey bands.



\section{Conclusions} 
In this paper, we present an alternative paradigm to infuse non-trivial topological characteristics into a trivial insulator using a scalar potential. The scalar potential is utilised to enhance the mixing between different quantum states which in turn drives the system into a topologically non-trivial regime accompanied by a reversal in \textit{mass} term. This switching is present in both one and two dimensions. In one dimension, it produces strong edge localisation whereas in two dimensions it shows the appearance of the chiral edge states with specific group velocity. In two dimensions, it gives rise to an emergent Hall effect which can be verified by calculating the Chern number.  In addition, our method also allows us to control the topological properties by local means, which is not possible with a conventional topological insulator. We demonstrate that the edge states can be controlled by selective placement of the scalar potential. 
One can observe the same qualitative behaviour with a periodic as well as non-periodic distribution of scalar potential as long as it does not form clusters. These predictions can be realised in real materials available experimentally. A suitable candidate for such a study would be a CdTe-HgTe-CdTe quantum well where the topological phases can be controlled by changing the width of the well. The scalar potential can be designed with suitable fabrication techniques \cite{Caro1986, Wang2020a} or can be introduced via a nonmagnetic dopant. For Hg$_{0.32}$Cd$_{0.68}$Te-HgTe quantum well, the mass gap ($2m$) is $\sim$50\,meV for a thickness of 50\,$\rm \AA$ \cite{Bernevig2006} which indicates the scalar potential induced topological transition can be observed for $V_0 \lesssim 0.5$\,eV. Our results thus open several new possibilities to control the topological properties and designing highly controllable devices for topological electronics.

\section{Acknowledgments} 
SG would like to acknowledge helpful discussions with Emil Prodan. We also gratefully acknowledge financial support by the Deutsche Forschungsgemeinschaft (DFG, German Research Foundation) - TRR 288/2 - 422213477 (project B06),  TRR 173/2 - 268565370 (project A11), CRC 1238 - 277146847 (Project C01), and the J\"ulich Supercomputing Centre for providing computational resources under the project jiff40.


\appendix

\section{Transmission of massive Dirac particle in one dimension through a rectangular barrier \label{KT}}

\begin{figure}[ht!]
\centering
\includegraphics[width=0.45\textwidth]{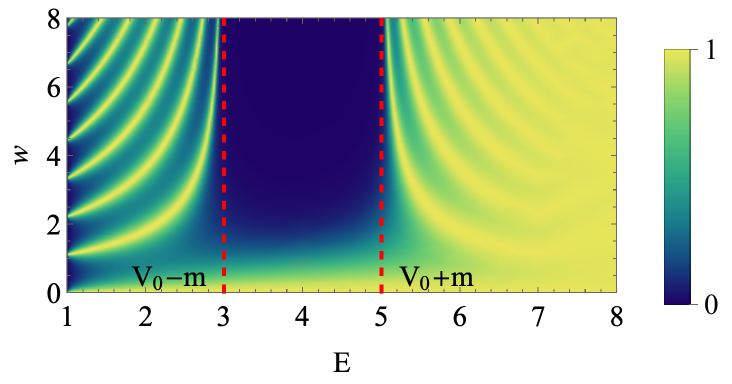}
\caption{Variation of transmission probability for different energy and barrier width in case of a rectangular barrier introduced in a one-dimensional  Dirac Hamiltonian. Here we choose $V_0$=4 and $m$=1.}
\label{fig:tr}
\end{figure}

Here we briefly show the transmission of a massive Dirac particle through a scalar potential which can provide a better understanding of the modulation of the mass term. We start with a massive Dirac equation in one dimension given by
\begin{eqnarray}
 H_1^D = -i \sigma_1 \partial_x + \sigma_3 m + \sigma_0 V(x).
 \label{H1a}
\end{eqnarray} 
where $m$ is the mass term which we choose to be 1. $\sigma_{1,3}$ are the Pauli matrices and $\sigma_0$ is the identity matrix of rank 2. The simplest way to study the impact of a scalar potential is to introduce a rectangular barrier of width $w$, such that $V(x)=V_0$ for $-w/2 \leq x \leq w/2$ and $V(x)=0$ otherwise. The transmission probability for such a rectangular barrier is  given by $T=16 \left\vert \lambda/ \left[ (1+\lambda)^2e^{-i\kappa_2 w}-(1-\lambda)^2e^{i \kappa_2 w} \right] \right\vert^2$  where $\kappa_1=\sqrt{E^2-m^2}$ and $\kappa_2=\sqrt{(E-V_0)^2-m^2}$ and $\lambda = \frac{\kappa_2}{\kappa_1}\frac{E+m}{E-V_0+m}$. For such a system, it is possible to achieve complete transmission probability if the barrier height is greater than twice the mass term ($V_0>2m$). which is manifested as multiple  transmission channels for $E<V_0$ in Fig.\,\ref{fig:tr}.

For $w \to \infty$, the \textit{Klein window} ($m \leq E \leq V_0-m$) can manifest complete transmission. At this limit, the region inside and outside the scalar potential is dominated by a particular sign of the mass term. For small $w$, there is more mixture between the states with a different mass term which manifests itself as a modulation of the transmission probability with respect to the energy on both sides of the forbidden zone ($V_0-m \leq E \leq V_0+m$). In a periodic lattice with a fixed width of the potential region, this interference results in a modulation of the mass gap with respect to the barrier height $V_0$.

\section{Random orientation of potential}

\begin{figure}[h!]
\centering
\includegraphics[width=0.48\textwidth]{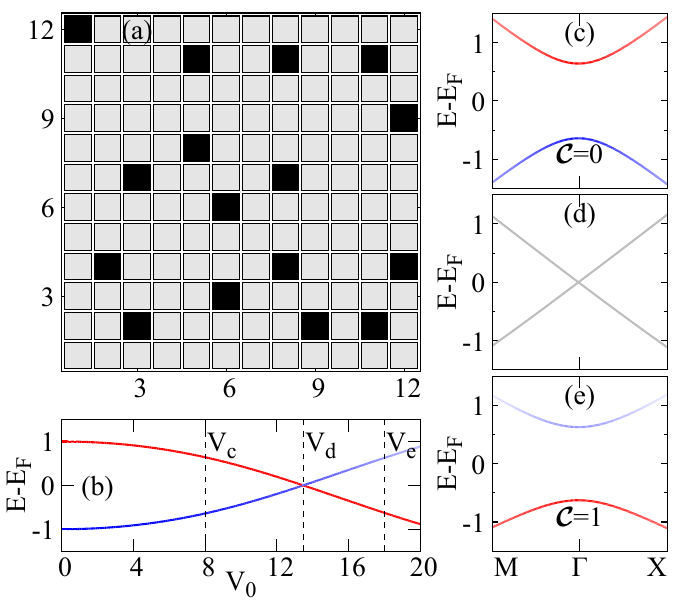}
\caption{Variation of mass gap and Chern number for supercell. (a) Distribution of potential (black blocks) over the supercell. (b) Variation of band gap where the red and blue colour denotes positive and negative values of $\langle \sigma_z \rangle$. (c),(d) and (e) show band structure and Chern number for three different values of $V_0$ denoted by the vertical dashed line in (b).}
\label{fig:rbulk}
\end{figure}

\begin{figure}[h!]
\centering
\includegraphics[width=0.48\textwidth]{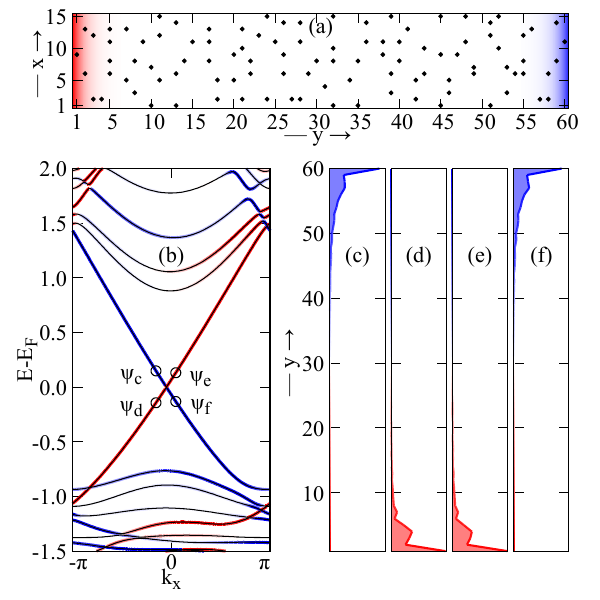}
\caption{Edge state formation for extended ribbon. (a) Distribution of potential over the ribbon. (b) The band structure of the ribbon where red and blue colour show the contribution from $y$=1 and $y$=60 edges. (c),(d),(e) and (f) show the average probability density along \textit{x} axis of four different states marked in (b).}
\label{fig:redge}
\end{figure}

The obvious question that arises at this point is whether this topological transition is an artefact of a perfect periodic system or not. To answer that we consider a $12 \times 12$ supercell with 16 scalar potential ($11.1\%$ coverage) and calculate the variation of mass gap (Fig.\,\ref{fig:rbulk}). Since calculating the Chern number for such a large system is computationally quite demanding, we calculate it for two different topological phases. One can readily see that the mass gap and the topological features change in a similar way as the small and uniform supercell (Fig.\,\ref{fig:band2}) which establishes that it is an intrinsic property and does not depend on the distribution of scalar potential.
Similar behaviour can be observed with a ribbon with open boundary condition along $y$ direction (Fig.\,\ref{fig:redge}). We choose a supercell of $15 \times 60$ sites and scatter 100 scalar potential with $V_0$=20. Here we also observe a pair of chiral edge states similar to the case of uniform distribution (Fig.\,\ref{fig:band2e}). Note that there is a small asymmetry in the bulk states which comes from the asymetric distribution of the scalar potential. However, it does not affect the presence of the edge states. These results confirm that the topological transition can be achieved with scalar potential for any arbitrary distribution.

\pagebreak

\bibliography{ref}

\end{document}